\documentclass{revtex4}

\usepackage{psfrag}
\usepackage{natbib}
\usepackage{amssymb}
\usepackage{amsmath}
\usepackage[dvips]{graphicx}
\usepackage{graphicx,epsfig}
\usepackage{latexsym}
\newcommand{\be}{\begin{equation}}
\newcommand{\ee}{\end{equation}}
\newcommand{\bi}{\begin{itemize}}
\newcommand{\ei}{\end{itemize}}
\newcommand{\ben}{\begin{enumerate}}
\newcommand{\een}{\end{enumerate}}

\newcommand{\ba}[1]{\overline{#1}}
\newcommand{\ffi}{\varphi}
\newcommand{\ffii}{\varphi_i}
\newcommand{\ffie}{\varphi_e}
\newcommand{\eps}{\varepsilon}
\newcommand{\Ai}{\mbox {Ai}}
\newcommand{\Bi}{\mbox {Bi}}
\newcommand{\Gi}{\mbox {Gi}}

\begin{document} % must come before the abstract

%%%%%%%%%%%%%%%%%%%%%%%%%%%%%%%%%%%%%%%%%%%%%%%%%%%%%%%%%%%%%%%%%

\title[FOLLOWER FORCE]{EDGE  FLUTTER OF  LONG  BEAMS UNDER FOLLOWER LOADS}

\keywords{Optional here; must be supplied on submission form}

%%%%%%%%%%%%%%%%%%%%%%%%%%%%%%%%%%%%%%%%%%%%%%%%%%%%%%%%%%%%%%%

\author{Emmanuel \surname{de Langre}}
\email{delangre@ladhyx.polytechnique.fr}
\affiliation{D\'epartement de M\'ecanique, LadHyX, Ecole Polytechnique, Palaiseau, France}
\author{Olivier \surname{Doar\'e}}
\email{olivier.doare@ensta-paristech.fr}
\affiliation{Unit\'e de M\'ecanique (UME), ENSTA-Paristech, Palaiseau, France}

%%%%%%%%%%%%%%%%%%%%%%%%%%%%%%%%%%%%%%%%%%%%%%%%%%%%%%%%%%%%%%%%%

\begin{abstract}
 The linear  instability of a beam tensioned by its own weight is considered. It is shown that for long  beams, in the sense of an adequate dimensionless parameter, the characteristics of the instability caused by a follower force do not depend on the length. The asymptotic regime significantly differs from that of  short beams: flutter prevails for all types of follower loads, and flutter is localized at the  edge of the beam. An approximate solution using matched assymptotic expansion is proposed for the case of a semi-infinite beam. Using a local criterion based on the stability of waves,  the characteristics of this regime as well as its range of application can be well predicted. These results are finally discussed in relation with cases of flow-induced instabilities of slender structures.   
\end{abstract}

\maketitle % must come after the abstract
\today

%%%%%%%%%%%%%%%%%%%%%%%%%%%%%%%%%%%%%%%%%%%%%%%%%%%%%%%%%%%%%%%%%

\section{Introduction}\label{intro}

The  linear stability of a beam under the action of a follower force exerted at one of its extremities has been the  subject of intensive research, as can be seen from the extensive review in  \cite{langthjem2000a}. The interest in this problem lies in its potential applications and also in the large variety of fundamental topics of mechanics involved in the solution method. Practical examples of direct application in the field of fluid-structure interactions are numerous: fluid-conveying pipes, plates subjected to  axial flow or towed cylindrical bodies are modelled by equations that are similar, though not identical, to those of a beam under a compressive follower force \citep{paidoussis1998a,paidoussis2003a}. More generally, follower forces have been extensivelly discussed in the literature, including in terms of their physical reality, see  \cite{elishakoff2005a} for a full  review. The case of a cantilevered beam of finite length under a partial follower force is well documented, see \cite{bolotin1963a}, with many results on the effect of characteristics of the beam or of the load on the critical load that causes  instability, and on the nature of the instability, be it divergence (buckling) or flutter. 

We seek here to establish the  characteristics  of  instability of a beam in the case where its length is much larger than the region where an unstable motion   will develop. This arises when a constant load, such as gravity acting on a vertically hanging beam, produces a tension that increases along the beam,  from zero at the lower free end to a maximum at the upper fixed end.  The increasing tension induces a corresponding increasing stiffness. Motion is then confined to the lower end, corresponding to {\em edge flutter}. This has been studied both experimentally and numerically in three  of the problems of flow-induced vibrations mentioned above: hanging fluid-conveying pipes \citep{delangre2001a,doare2002a}, hanging ribbons under axial flow \citep{lemaitre2005a} and towed cylinders under axial flow \citep{delangre2007a,obligado2013a}. In all these systems it was observed  that there exists a limit state in which  the length does not affect the stability. This limit state is found for length larger  than a limit value given by simple considerations on the local stability of bending waves  \citep{doare2002a,delangre2007a}. 

The objective of this paper is to establish similar results on the generic case of a beam under a partially follower force (or subtangential force), tensioned by a load such as gravity.  

In Section \ref{sec:equations} we shall recall the equations of motion and the possible choices of dimensionless variables. The effect of the beam length on  stability is analysed in Section \ref{sec:finite}, using numerical computations of the eigenmodes. In Section \label{sec:semiinfinite} we address the particular case of a semi-infinite beam, using various types of modelling. The application of the  results given in the paper to problems of flow-induced instabilities is discussed in Section \ref{sec:discussion}. 

%%%%%%%%%%%%%%%%%%%%%%%%%%%%%%%%%%%%%%%%%%%%%%%%%%%%%%%%%%%%%%%%%

\section{\label{sec:equations}Equations of motion}
\begin{figure}[h]
\begin{center}
\psfrag{psy}{$Y(Z,T)$}
\psfrag{psz}{$Z$}
\psfrag{psg}{$g$}
\psfrag{pspnf}{$(1-\eta)P$}
\psfrag{pspf}{$\eta P$}
\includegraphics[width=6cm]{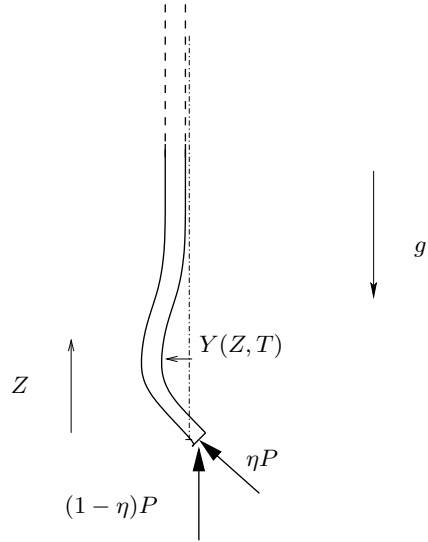}
\caption{\label{figpoutre} Semi-infinite hanging beam with a partially follower force. }
\end{center}
\end{figure}

We consider a vertical beam of length $L$ loaded by its own weight, Fig.\ref{figpoutre}.  

A partial follower force is applied at its lower end \citep{langthjem2000a}. The linear equation governing the in-plane lateral deflection  $Y(Z,T)$  reads 
\be
\label{equlibredim}
EI \frac{\partial^4 Y}{\partial Z^4}+ \frac{\partial}{\partial Z}\left[(P-mgZ)\frac{\partial Y}{\partial Z}\right]+ m\frac{\partial^2Y}{\partial T^2}=0,
\ee
where $EI$ is the flexural rigidity, $P$ is the load, $g$ is gravity and  $m$ is the mass per unit length. No damping is considered here, though it is known to significantly influence some aspects of the problem, see for instance in \citep{detinko2003a}. 
The boundary conditions at the lower  end, $Z=0$ are 
\be
EI \frac{\partial^2Y}{\partial Z^2}(0)=0\  ,\  EI \frac{\partial^3Y}{\partial Z^3}(0)+ (1-\eta) P \frac{\partial Y}{\partial Z}(0)=0 ,
\ee
where $\eta$ is a coefficient that  expresses the part of the loading that follows the beam slope; hence $\eta=0$  corresponds to   a non-follower force and  $\eta=1$ to a pure follower force. At the upper end, we assume a clamped condition,
\be
Y(L)=0, \ \frac{\partial Y}{\partial Z}(L)=0.
\ee

Using the length of the beam, $L$, as a  reference we define the following dimensionless variables:
\be
z=\frac{Z}{L}, \ y=\frac{Y}{L},\  t= \left(\frac{EI}{m}\right)^{1/2} \frac{1}{L^2}T, \  p=\frac{PL^2}{EI}, \gamma=\frac{mgL^3}{EI}.
\ee
Then Eq.(\ref{equlibredim}) may be re-written in dimensionless form as 
\be
\label{equsuivadim}
 \frac{\partial^4 y}{\partial z^4}+ \frac{\partial}{\partial z}\left[(p-\gamma z)\frac{\partial y}{\partial z}\right]+ \frac{\partial^2 y}{\partial t^2}=0,
 \ee
with the boundary conditions
\be
\label{equbc0}
 \frac{\partial^2 y}{\partial z^2}(0)=0  ;  \  \frac{\partial^3 y}{\partial z^3}(0)+(1-\eta) p\frac{\partial y}{\partial z}(0)=0; \ y(1)=0;\ \frac{\partial y}{\partial z}(1)=0.
 \ee
This dimensionless set of equations is adequate to analyse the effect of $\gamma$ on the critical loading, but only if the length $L$ is kept constant. As we need to vary the length  $L$, it is necessary to define a new set of dimensionless parameters.  We  now use  a length scale defined by  the ratio of the two stiffnesses of the beam,  namely the flexural rigidity and the stiffness related to tension induced by gravity \citep{doare2002a}, 
\be
\label{equLg}
L_g= \left(\frac{EI}{mg}\right)^{1/3}
\ee
which will be referred to as the gravity length. Note that this length scales like the critical length that causes buckling of the beam under its own weight.  Using $L_g$ as the reference length to define  dimensionless variables we have  
\be
\label{equvardim2}
x=\frac{Z}{L_g},\  y=\frac{Y}{L_g},\  \tau= \left(\frac{EI}{m}\right)^{1/2}\frac{1}{L_g^2}T,\ q=\frac{PL_g^2}{EI}, \  \ell=\frac{L}{L_g},
\ee
 where $L_g$ has been substituted for $L$.
These are related to the previous set of variables by 
 \be
x=z\ell,\  \tau=t \ell^2,q =\frac{p}{\ell^2}, \ \ell=\gamma^{1/3}.
\ee
Eq. (\ref{equlibredim})  now becomes
\be
\label{equsuivadim2}
 \frac{\partial^4 y}{\partial x^4}+ \frac{\partial}{\partial x}\left[(q-x)\frac{\partial y}{\partial x}\right]+ \frac{\partial^2 y}{\partial \tau^2}=0,
 \ee
and the corresponding boundary conditions are
\be
\label{equbc1}
 \frac{\partial^2 y}{\partial x^2}(0)=0  ;  \  \frac{\partial^3 y}{\partial x^3}(0)+(1-\eta) q\frac{\partial y}{\partial x}(0)=0; \ y(\ell)=0;\ \frac{\partial y}{\partial x}(\ell)=0.
 \ee

%%%%%%%%%%%%%%%%%%%%%%%%%%%%%%%%%%%%%%%%%%%%%%%%%%%%%%%%%%%%%%%%%

\section{\label{sec:finite}Stability of finite beams}

%%%%%%%%%%%%%%%%%%%%%%

\subsection{Solution for short beams}

\begin{figure}
\begin{center}
\psfrag{pseta}{$\eta$}
\psfrag{psomr}{$\Omega_c$}
\psfrag{pspc}{$p_c$}
\psfrag{psa}{(a)}
\psfrag{psb}{ (b)}
 \centerline{
\includegraphics[width=8cm]{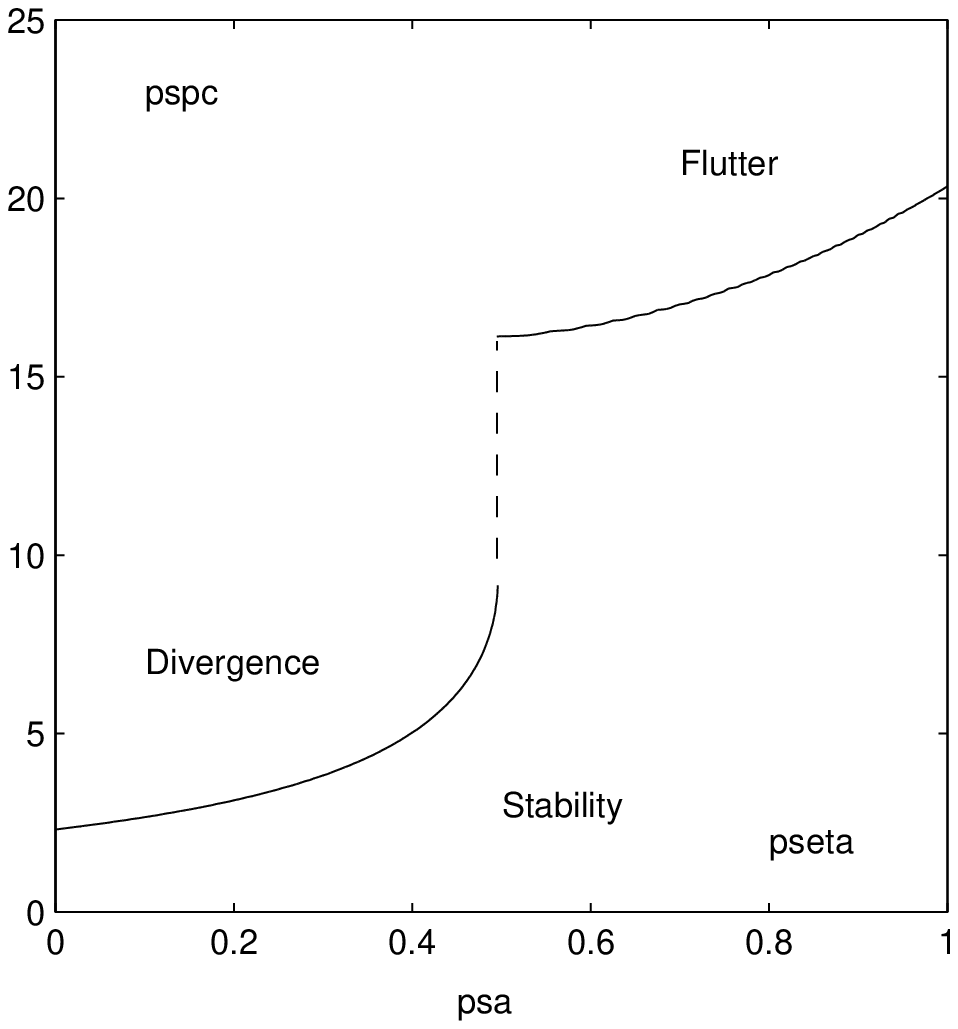}
\includegraphics[width=8cm]{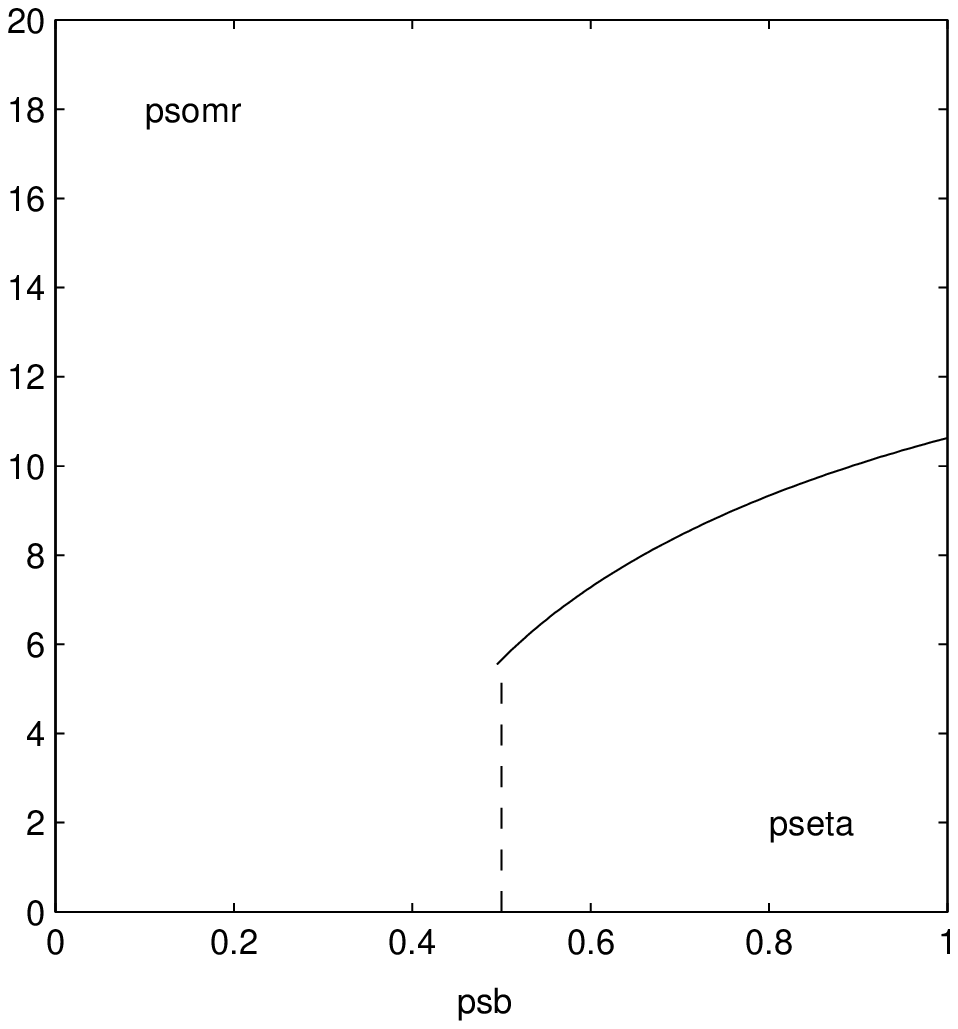}}
\caption{\label{figvalidationcourt}Critical load, (a), and frequency at  instability, (b), for short beams. Dashed vertical lines are guides for the eyes that do not correspond to calculations. }
\end{center}
\end{figure}

For short beams, in the sense where $\ell \ll 1$, the critical load $q_c$ may be derived by considering  that  $\gamma \ll 1$ so that Eq. (\ref {equsuivadim}) becomes 
\be
\label{equsuivadimshort}
 \frac{\partial^4 y}{\partial z}+ p\frac{\partial^2 y}{\partial z^2}+ \frac{\partial^2 y}{\partial t^2}=0.
 \ee
The corresponding stability diagram of $p_c$ versus $\eta$ is that of the generalized Beck's column \citep{langthjem2000a} and is shown in Fig.\ref{figvalidationcourt}(a). For $\eta<0.5$ instability arises in the form of a divergence, whereas flutter prevails for $\eta>0.5$.
The corresponding frequency at the flutter limit is shown in Fig.\ref{figvalidationcourt}(b).  
The critical load  and frequency at flutter expressed in   the dimensionless variables  of Eq. (\ref{equvardim2}) are 
\be
\label{equshort}
q_c(\eta,\ell)=\frac{p_c(\eta)}{\ell^2}; \  \omega_c(\eta,\ell)=\frac{\Omega_c(\eta)}{\ell^2}.
\ee

%%%%%%%%%%%%%%%%%%%%%%

\subsection{\label{sec:lengtheffect}Effect of length}

We now investigate the effect of the length $\ell$ on  the stability threshold and the nature of the instability. 
The numerical method  utilized to this end is the finite difference method,  as in \citep{Sugiyama1975a}. It allows us  to compute the eigenvalues $\omega$ and  the eigenvectors $\ffi$, such that  
\be
\label{equfifi}
\ffi^{(4)}+ [(q- x)\ffi']'-\omega^2 \ffi =0
 \ee
with 
\be
\ffi''(0)=\ffi^{(3)}(0)+(1-\eta) q \ffi'(0)=\ffi(\ell)=\ffi'(\ell)=0.
\ee 
The critical load and the nature of the  corresponding instability can be determined from the evolution of the real and imaginary parts of the frequency.  
The beam is unstable if the frequency $\omega$ has a negative imaginary part.
If the real part is zero, the instability results in an exponential growth in time of the deformation , without oscillation and the instability is of the divergence type. If the real part is non zero, the instability results in an exponential growth of oscillations, and the instability is of the flutter type.

\begin{figure}
\begin{center}
\psfrag{psl}{$\ell$}
\psfrag{psqc}{$q_c$}
\psfrag{psa}{}
\psfrag{psb}{}
\begin{tabular}{cc}
\includegraphics[width=8cm]{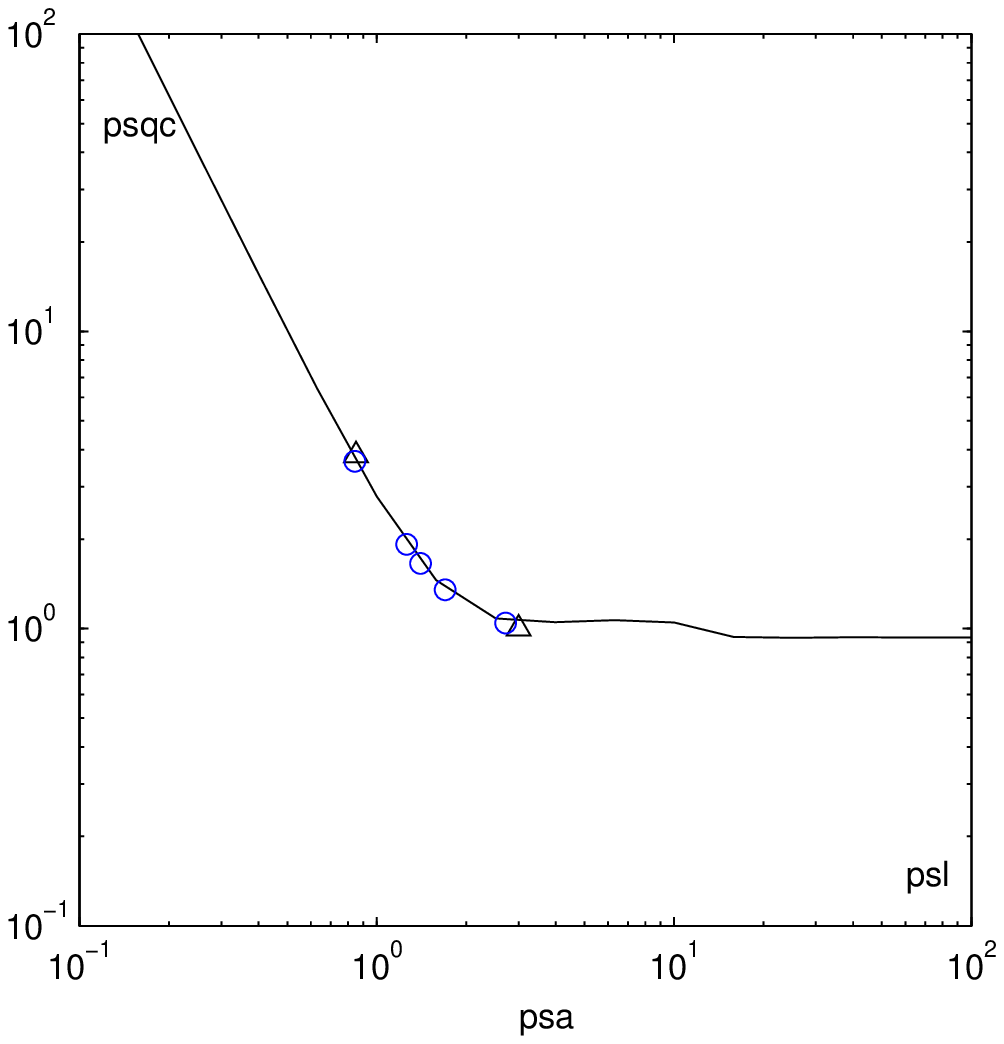}
& \includegraphics[width=8cm]{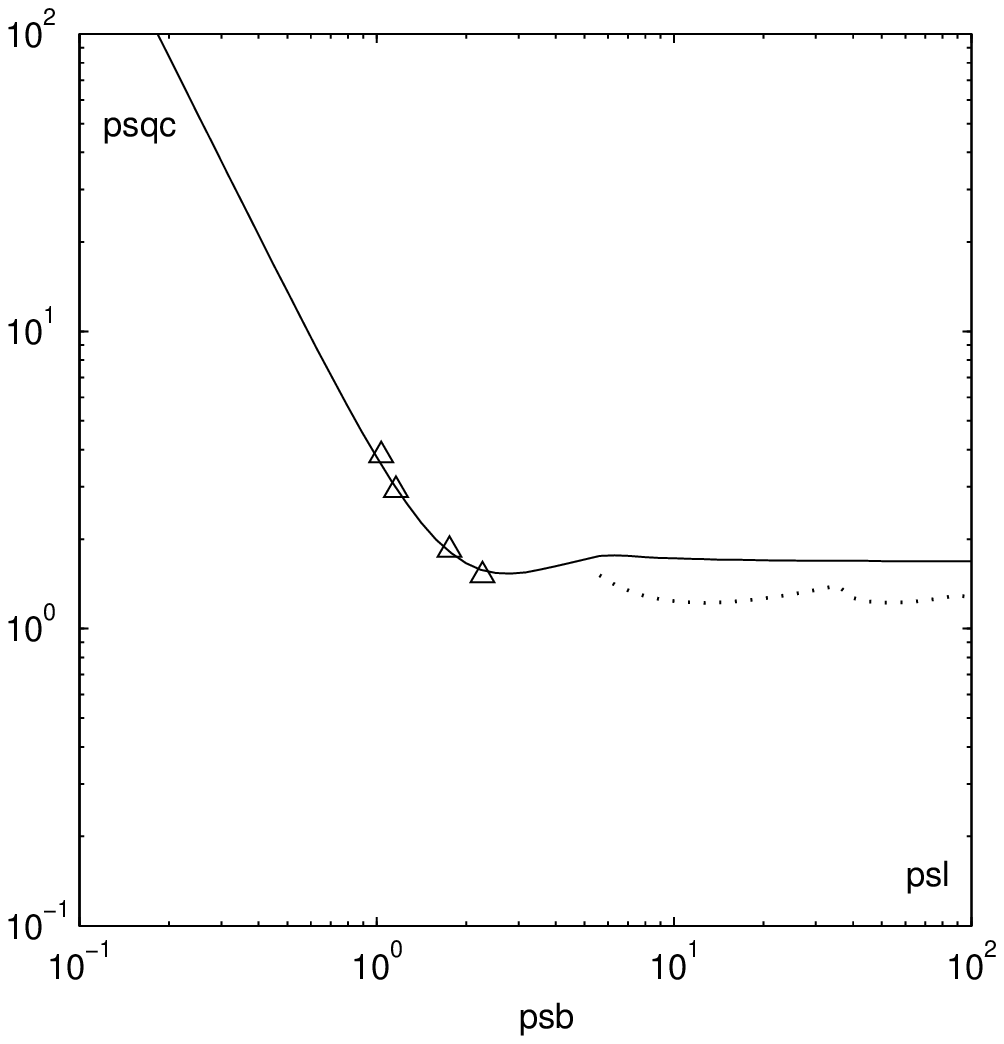} \\
(a) & (b) \\
\includegraphics[width=8cm]{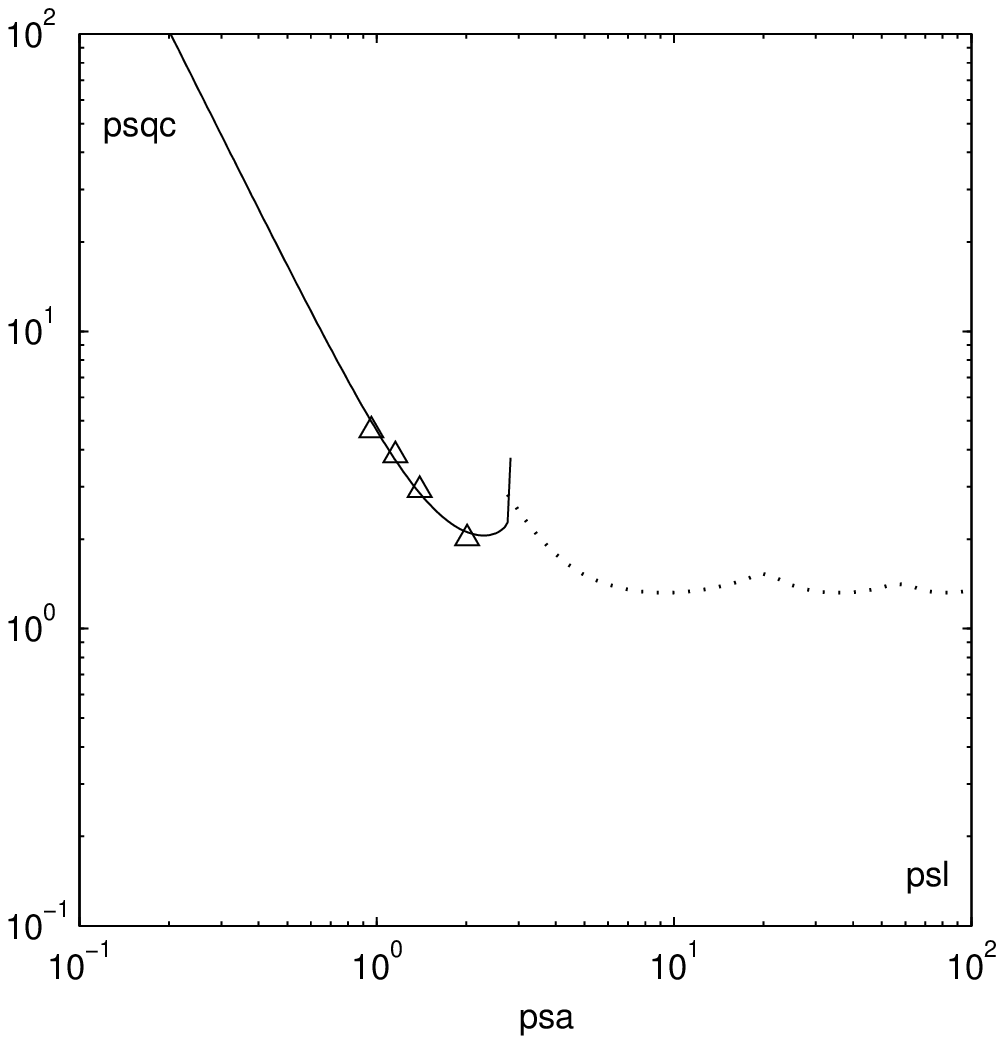}
& \includegraphics[width=8cm]{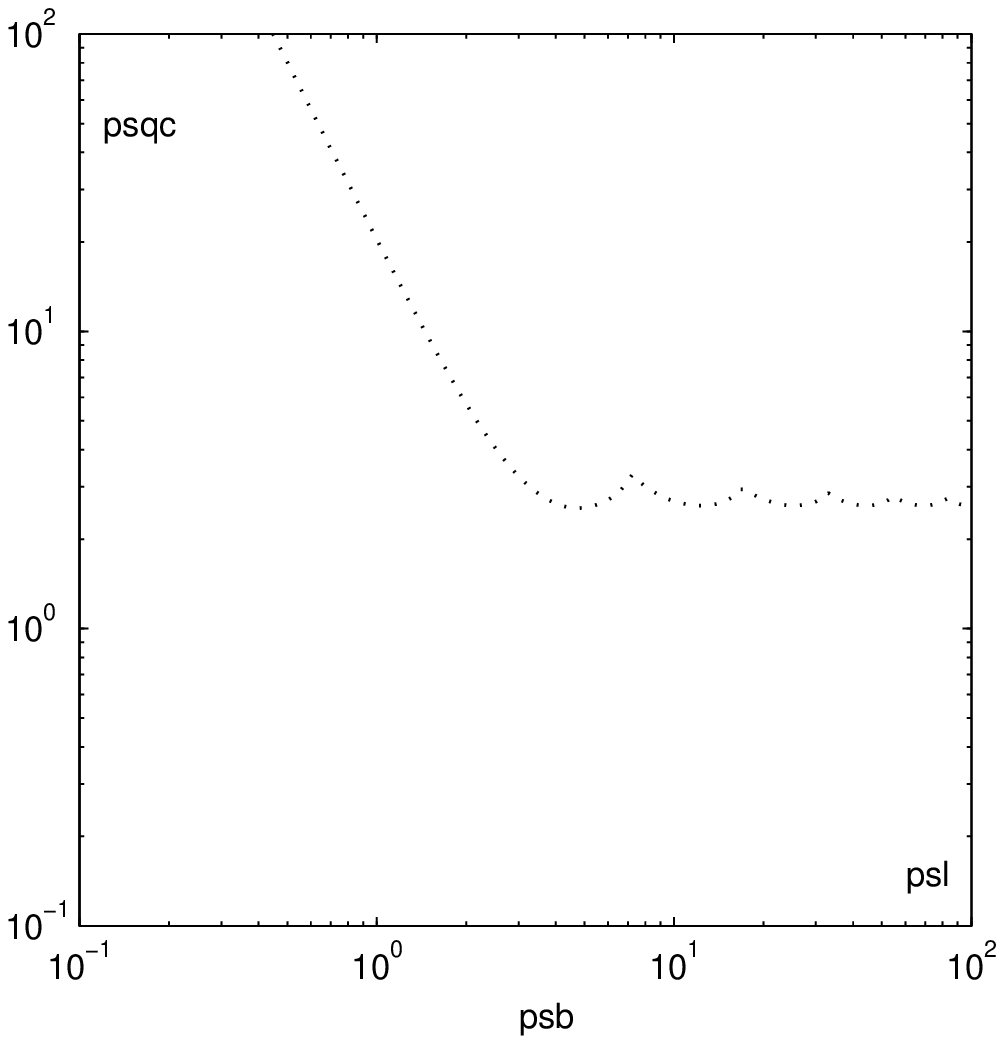} \\
(c) & (d)
\end{tabular}
\caption{\label{figeffetlongueur}Effect of length on the critical load for divergence and for flutter instabilities, (a) $\eta=0$,(b) $\eta=0.2$, (c) $\eta=0.3$, (d) $\eta=1.$. (---), present results for divergence; (- -), present results for flutter; (o), computations by \cite{Naguleswaran2004a} for $\eta=0$; ($\Delta$), computations by \cite{triantafyllou1984a}.}. 
\end{center}
\end{figure}

Fig.\ref{figeffetlongueur} show the evolution of the critical load with length, for several values of $\eta$ of particular interest.

For $\eta=0$, Fig.\ref{figeffetlongueur}(a), which is the case of a non-follower force, the critical load decreases steeply with length, up to about $\ell=5$ where it reaches a limit value and does not change when the length is increased further. Instability is for all lengths of the divergence type. On the same figure  are shown computations from  \cite{Naguleswaran2004a} and \cite{triantafyllou1984a}, for intermediate values of the length.  In \cite{triantafyllou1984a}, the static instability of a towed beam has been considered, which yields equations similar to that used here. Details of the equivalence are discussed in the last section of this paper, but suffice to say  here that we may use the results of Fig. 4 of \cite{triantafyllou1984a} with the change of variables $\ell=\varepsilon\lambda^{2/3}$ and   $q=\lambda^{2/3}$. It is seen that their results and those of this paper are in very good agreement. 
For $\eta=0.2$, Fig.\ref{figeffetlongueur}(b), the critical load for divergence also decreases and reaches a limit value, but the flutter threshold becomes lower than that for divergence, for   beams longer than  $\ell = 5$.
For $\eta=0.3$, Fig.\ref{figeffetlongueur}(c),  no divergence is found for $\ell>3 $. Instability is that range of $\ell$ is  of the flutter type. 
For $\eta=1$, Fig.\ref{figeffetlongueur}(d), the system loses stability by flutter, similarly to Beck's column; the critical load for flutter decreases with $\ell$ until it reaches a limit value.

From these four cases it can be stated that there exists a limit configuration for all long beams,
say $\ell > 10 $. For these lengths, the type of instability, flutter or divergence, may differ from that observed for short beams at the same value of $\eta$.

\begin{figure}[htb]
\begin{center}
\psfrag{pseta}{$\eta$}
\psfrag{psqc}{$q_c$}
\psfrag{psa}{(a)}
\psfrag{psb}{(b)}
 \centerline{
\includegraphics[width=8cm]{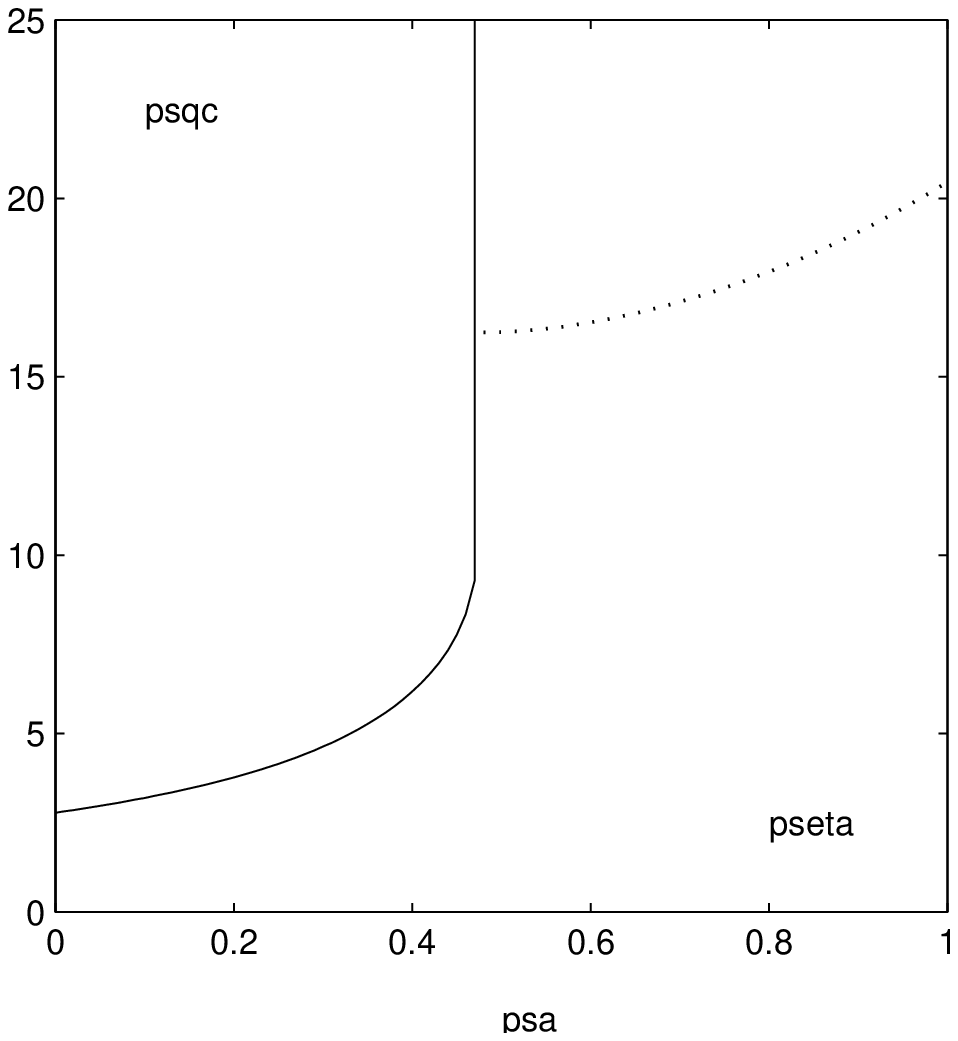}
\includegraphics[width=8cm]{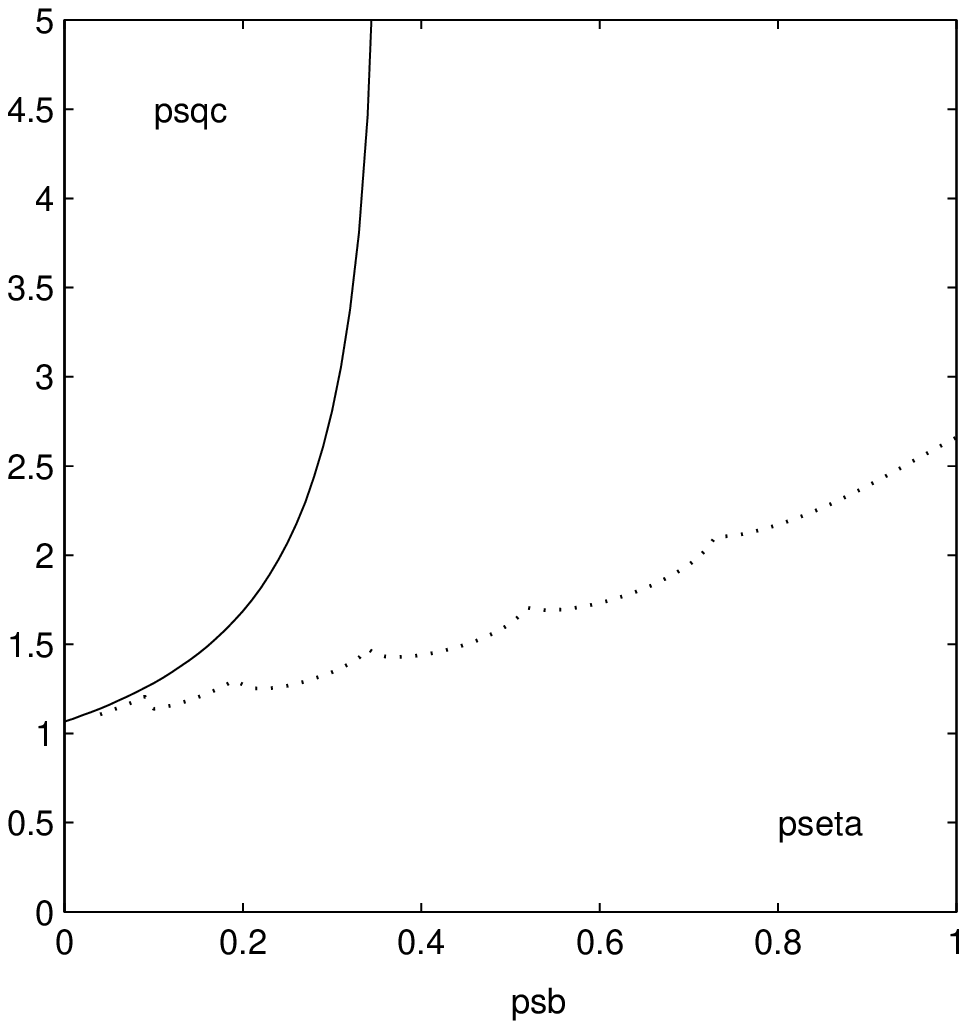}}
\caption{\label{figevoldiagramme} Effect of the  ratio $\eta$ on the critical load for divergence, (---), and for flutter, (- -). (a) Short beam, $\ell=1$, (b) Long beam,  $\ell=100$.} 
\end{center}
\end{figure}

This is further confirmed in the evolutions of the stability diagram $q_c(\eta)$ shown in Fig.\ref{figevoldiagramme}. At $\ell=1$ the diagram is similar, though not identical, to that for short beams; since  $\ell=1$  we have  here $q_c=p_c$  and   Figs. \ref{figvalidationcourt} and \ref{figevoldiagramme}(a) can be directly compared.
At $\ell=100$, Fig.\ref{figevoldiagramme}(b) a limit state is  almost reached, where stability is always lost by flutter, for all values of $\eta$. 

\begin{figure}
\begin{center}
\psfrag{psfi}{$\ \ \varphi$}
\psfrag{psz}{$x$}
\psfrag{psa}{(a)}
\psfrag{psb}{(b)}
\psfrag{psc}{(c)}
\includegraphics[width=11cm]{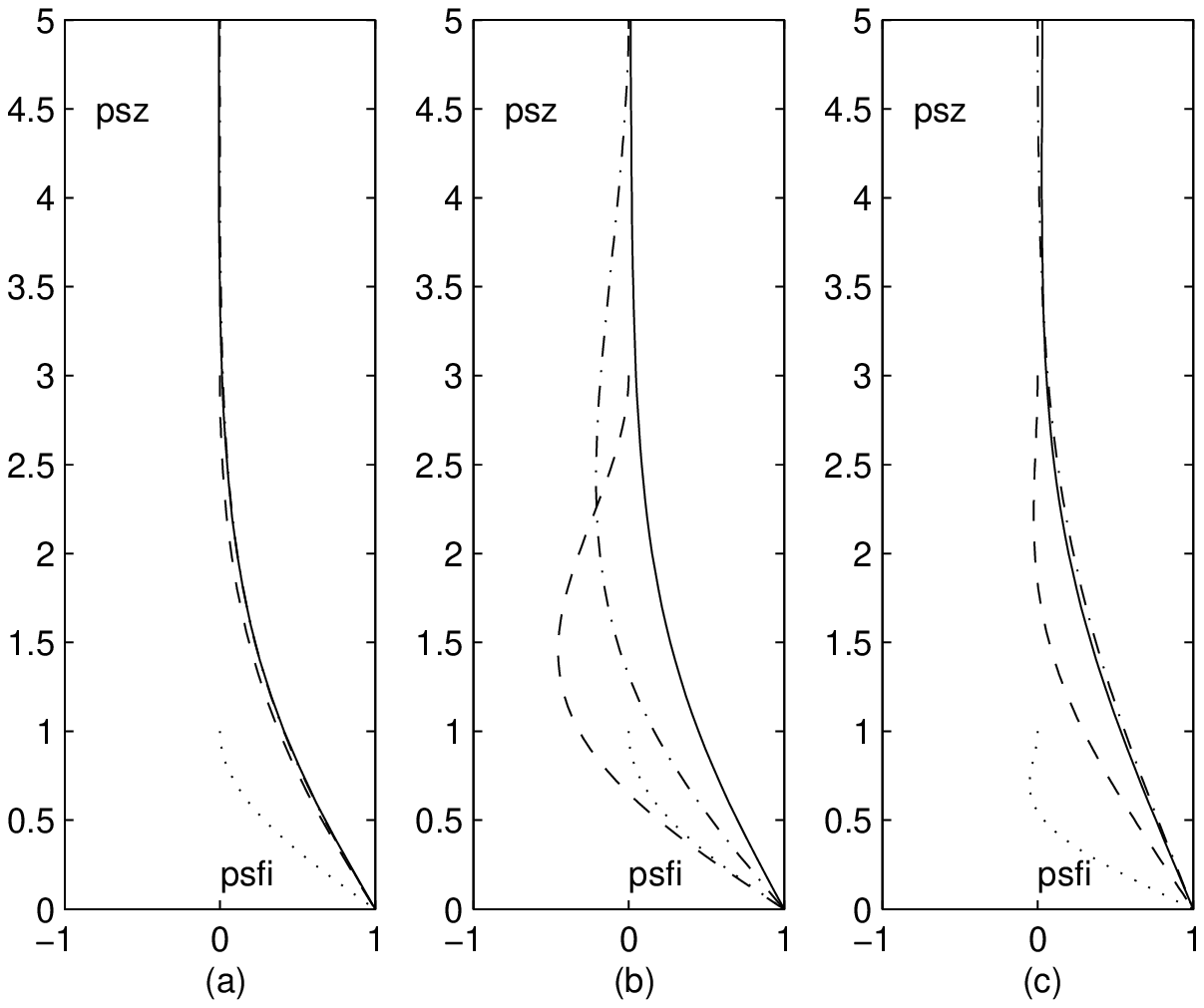}
\caption{\label{figdeftot} Mode shape at the critical load. (a) $\eta=0$, (b) $\eta=0.3$, (c) $\eta=1$. (..), $\ell=1$, (- -), $\ell=3$; (-.-), $ \ell=5$; (---), $ \ell=10$.}
\end{center}
\end{figure}

Fig.\ref{figdeftot} shows the mode shape at the   critical load, which may be a divergence or a flutter instability depending on the values of $\eta$ and $\ell$, as shown above. Note that at the instability threshold $\varphi$ is real, even for flutter instability. In fact, in Eq. (\ref{equfifi}), when $\omega$ is real so is $\varphi$.  For $\eta=0$, Fig.\ref{figdeftot}(a), the mode shape, here for divergence, becomes independent of the length when $\ell >3$. This is consistent with Fig.\ref{figeffetlongueur}(a) where the critical load was found to be stationnary in this range.
For $\eta=0.3$ and $\eta=1$, Figs. \ref{figdeftot}(b) and (c), the mode shapes also converge to a constant shape as the length is increased. There, the displacement is confined to the lower part of the beam. Note that the instability is of the flutter  type, except for $\eta=0.3$ at $\ell=1$. Yet, the limit mode shape closely resembles that of the divergence instability at $\eta=0$.  

%%%%%%%%%%%%%%%%%%%%%%%%%%%%%%%%%%%%%%%%%%%%%%%%%%%%%%%%%%%%%%%%%

\section{\label{sec:semiinfinite}Stability  of a semi-infinite beam}

%%%%%%%%%%%%%%%%%%%%%%

\subsection{Boundary conditions}

We now seek to determine the characteristics of the instability when the beam is assumed to be of infinite length in the $X$ axis. This is expected to be the solution to which the results of the preceding section converge as $\ell$ is increased. 
Whereas the boundary condition at the lower end is unchanged, the condition that the beam is clamped at $X=L$ needs to be replaced by the conditions that the displacement  remains finite as $X$ goes to infinity and that in this limit propagating waves only radiate in the direction of increasing $X$. These conditions in dimensionless form read, 
\be
\label{equbcinf}
 \lim_{x \rightarrow \infty} |y| = 0  ;\  \lim_{x \rightarrow \infty} \frac{\partial y}{\partial x}\frac{\partial y}{\partial t} <0.
 \ee
 
%%%%%%%%%%%%%%%%%%%%%%
 
\subsection{Divergence instability}

The case of divergence instability may be analysed by neglecting all time derivatives in the equations,  so that (\ref{equsuivadim2}) becomes
\be
\label{equstat}
 \frac{\partial^4 y}{\partial x^4}+ \frac{\partial}{\partial x}\left[(q-x)\frac{\partial y}{\partial x}\right]=0
\ee
and the boundary conditions are  only
\be
\frac{\partial^2 y}{\partial x^2}(0)=0  ;  \  \frac{\partial^3 y}{\partial x^3}(0)+(1-\eta) q\frac{\partial y}{\partial x}(0)=0; \lim_{x \rightarrow \infty} |y| = 0. 
\ee
 By defining  $\lambda =q^{3/2}$ and $Z=x-q$, this set of equations is identical to that solved by \cite{triantafyllou1984a} for the case of a towed cylinder. Using their solution with our variables, the critical load for divergence $q_c(\eta)$ is found to satisfy
\be
\label{equt84}
\Ai'(-q_c)-q_c\eta \left[\frac{1}{3}-\int_{0}^{-q_c} \Ai(s)ds\right] =0,
\ee
where $\Ai$ is the Airy function \citep{abramowitz1970a}.

%%%%%%%%%%%%%%%%%%%%%%

\subsection{Flutter instability }

We give here an approximate solution for the critical load for flutter of a semi-infinite beam, which is an  extension  from the work of \cite{delangre2001a} for  a purely follower force, $\eta=1$. The approach used in the matched assymptotic expansions follows that proposed  by  \cite{triantafyllou1991a} for the dynamics of vibrating strings tensionned by gravity. 

We consider Eq. (\ref{equsuivadim2}), with the boundary conditions (\ref{equbc1}) at the bottom end, and  (\ref{equbcinf}) for a semi-infinite beam.
We seek harmonic solutions of this set of equations, of the form
\be
\label{equyharm}
y(x,t)= \mbox {Real}[\ffi(x) e^{i\omega t}]
 \ee
as a function of the follower force magnitude $q$. Instability will be associated with $\mbox {Im}(\omega)<0$.
The variation in space of the solution satisfies
 \be
\label{equfi}
\ffi^{(4)}+ [(q-x)\ffi']'-\omega^2  \ffi =0,
 \ee
with the boundary conditions
 \be
\label{equbcfi}
 \ffi^{(2)}(0)=0  ; \ffi^{(3)}(0)+(1-\eta)q\ffi'(0)=0  
 \ee
where $()'$ denotes derivation with respect to $x$. The condition for radiating waves reads
\be
\label{equbcfi2}
\lim_{x \rightarrow \infty} \left(\mbox {Real}[\ffi'(x) e^{i\omega t}]\mbox {Real}[i\omega \ffi(x) e^{i\omega t}]\right)<0.
\ee

Let us consider first the lower part of the beam where $x$ is of the order of $q$. In this ``inner'' domain of length $q$, we may derive an approximate solution $\ffii(x)$ by simply lumping all its inertia at the lower end of the beam, so that the boundary conditions  at $x=0$ read   now 
 \be
\label{equbcfi0}
 \ffii^{(2)}(0)=0  ; \ffii^{(3)}(0)+(1-\eta)q\ffii'(0)=q\omega^2 \ffii(0),
 \ee
and Eq.  (\ref{equfi}) reduces to
\be
\label{equfii}
\ffii^{(4)}+ [(q-x)\ffii']' =0.
 \ee
The corresponding  solution reads
\be
\label{equfii1}
\ffii(x)=a + \int_0^x [ b\Ai(s-q) + c\Bi(s-q) + d\Gi(s-q)]\  ds
 \ee
where $\Ai$, $\Bi$ and $\Gi$  are the Airy functions, $a,\ b,\ c, \ d$ being four coefficients. 
For the sake of clarity we now use the  notation $\ba{F}(x)=\int_{0}^x F(s)ds $ and $F$ for $F(-q)$ unless otherwise specified, where $F$ could be any of the Airy functions above.  

In terms of the coefficients  $a,\ b,\ c, \ d$  the boundary conditions  at the lower end are 
\be
\label{equfiibc}
b \Ai'+ c \Bi'+d \Gi'=0; 
\ee
\be
\label{equfiibc1}
  b \Ai''+  c \Bi''+d \Gi''+(1-\eta)q(b \Ai+  c \Bi+d \Gi)=  a q\omega^2. 
\ee

Conversely, in the upper part of the beam, that  is where $x\gg q$, we rescale the equations by using new variables in space and time, namely 
 \be
\label{equvarfee}
 \chi=x\eps;  \ r=\frac{\omega}{\sqrt{\eps}},
 \ee
where $\eps= L_g/\Lambda$, the length $\Lambda$ being an arbitrary large scale. Using this set of variables, Eq.  (\ref{equfi}) reads 
 \be
\label{equfieps}
\eps^3\ffi^{(4)}+ [(\eps^2 q-\eps \chi)\ffi']'-\eps r^2  \ffi =0,
 \ee
where $()'$ denotes now differentiation  with respect to $\chi$. This allows us  to derive the equation at the leading order in the ``outer''domain,  
\be
\label{equfie0}
(- \chi\ffie')'- r^2  \ffie =0,
 \ee
the solution of which, in terms of Bessel functions, is (see also in \citep{elishakoff1987a})
\be
\label{equfie}
\ffie(\chi)=\alpha J_0 ( 2r\sqrt{\chi})  +\beta Y_0 ( 2  r\sqrt{\chi}).
 \ee
The radiation condition, Eq. (\ref{equbcfi2}), 
implies  that 
\be
\label{equalfabeta}
\alpha-i\beta=0. 
 \ee

We may now match the inner and outer solutions by considering their respective limits \citep{abramowitz1970a,triantafyllou1991a}.
Equating these two limits yields three conditions for  the coefficients defining the inner and outer solutions, namely
\be
\label{equracc123}
c=0; \ d=\beta; \ a + b \left[ \frac{1}{3}- \ba{\Ai}\right]+d\left[\frac{2\gamma + \ln 3 }{3\pi}- \ba{\Gi}  \right]=\alpha + \beta \frac{2}{\pi} \left(\ln \omega + \gamma\right)
 \ee
where $\gamma$ is Euler's constant.
The set of boundary conditions (\ref{equfiibc}), (\ref{equfiibc1}), (\ref{equalfabeta}), with the three  matching conditions  (\ref{equracc123}) are six linear equations between the six coefficients $\alpha,\ \beta,\  a,\ b,\ c,\ d$. This defines an implicit relationship between the two parameters $\omega$ and $q$, so that there exists a non-trivial solution. It reads
\be 
\label{equomegap} 
-\eta q[\Gi'\Ai-\Gi \Ai']-\frac{1}{\pi}\Ai'+q\omega^2 \left[  \left( \frac{1}{3}-\ba{\Ai} \right)\Gi'-  \left( \frac{\ln 3-4\gamma -3i\pi-6\ln \omega }{3\pi} -\ba{\Gi}\right)\Ai'\right]=0.
 \ee

\begin{figure}
\begin{center}
\psfrag{psomr}{$\omega_c$}
\psfrag{pseta}{$\eta$}
\psfrag{psqc}{$q_c$}
\psfrag{psa}{(a)}
\psfrag{psb}{(b)}
 \centerline{
\includegraphics[width=8cm]{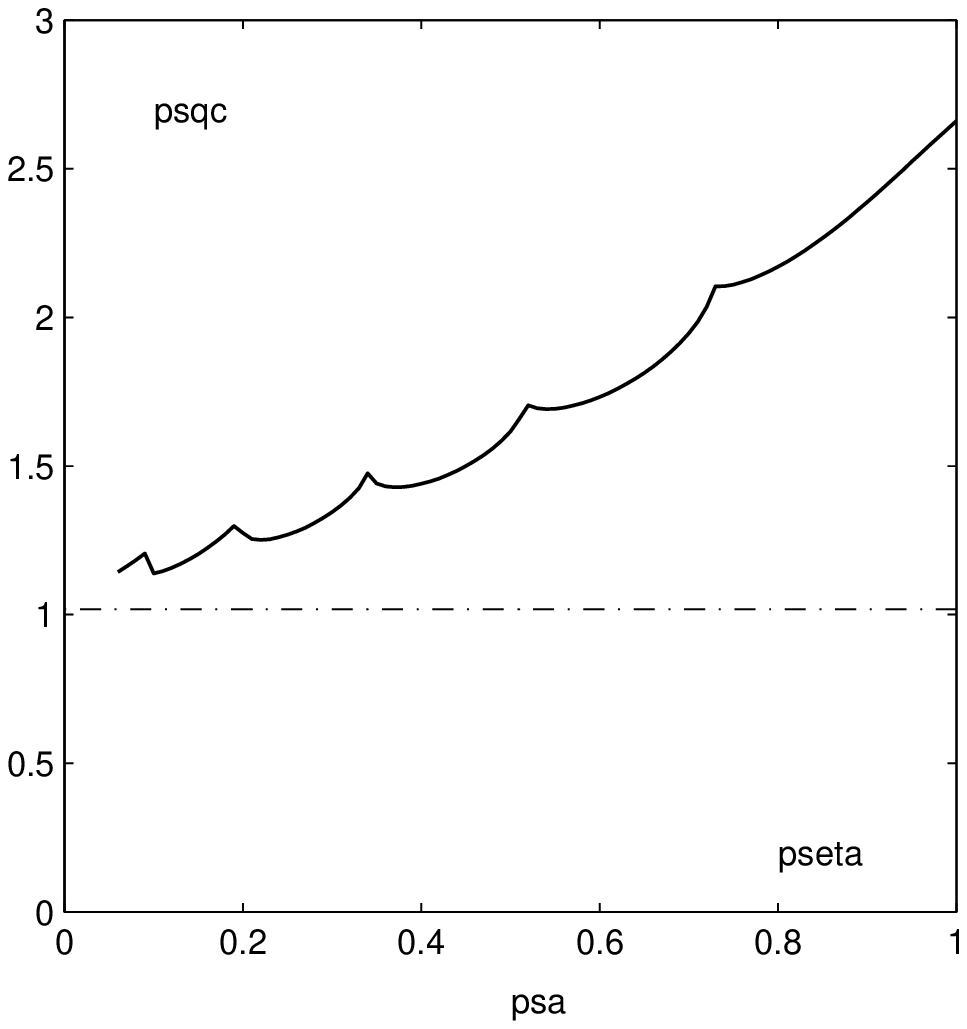}
\includegraphics[width=8cm]{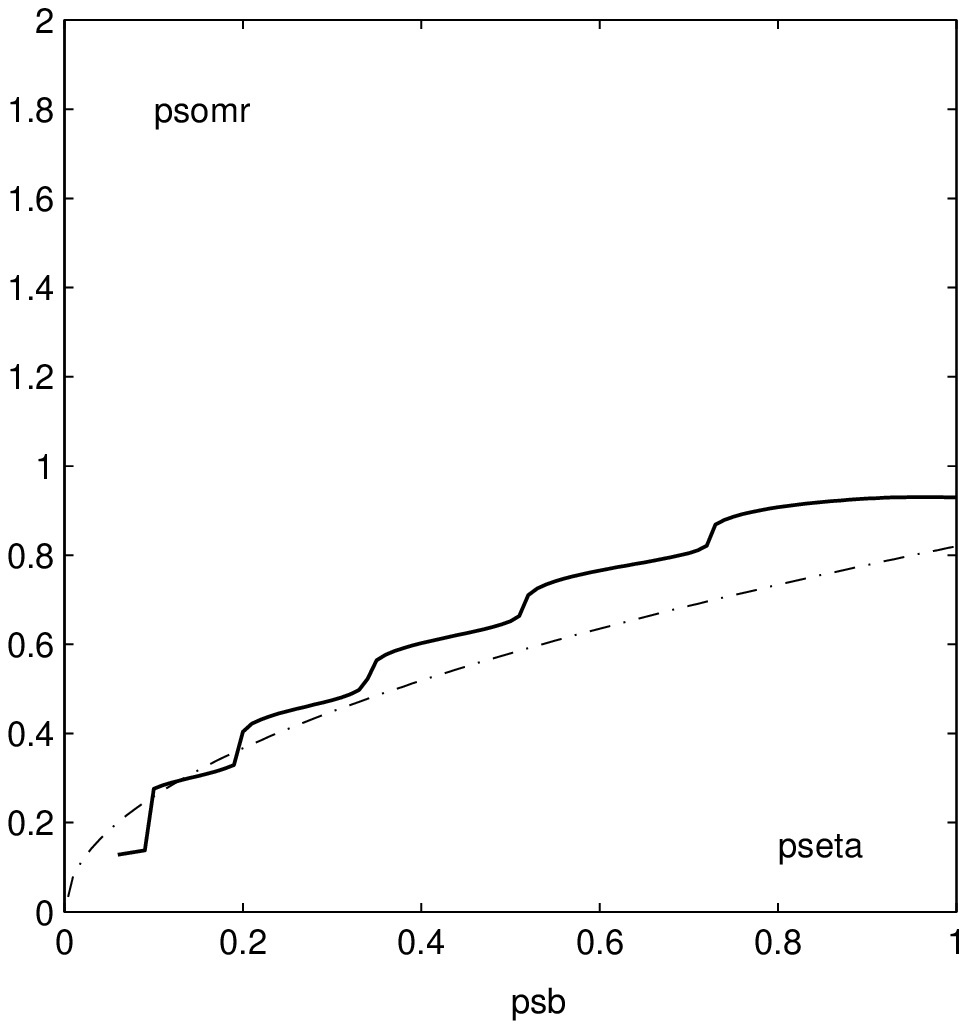}}
\caption{\label{figapproxdar}Critical load $q_c$, (a), and frequency $\omega_c$ of flutter,(b), of a long beam. \mbox{(---)}, numerical results for $\ell=100$; \mbox{(-.-)}  solution for the semi-infinite beam  using matched assymptotic expansions, Eqs. (\ref{equpc}) and  (\ref{equomegac}).}
\end{center}
\end{figure}

\begin{figure}
\begin{center}
\psfrag{psfi}{$\varphi$}
\psfrag{psz}{$x$}
\psfrag{psa}{(a)}
\psfrag{psb}{(b)}
\psfrag{psc}{(c)}
\includegraphics[width=11cm]{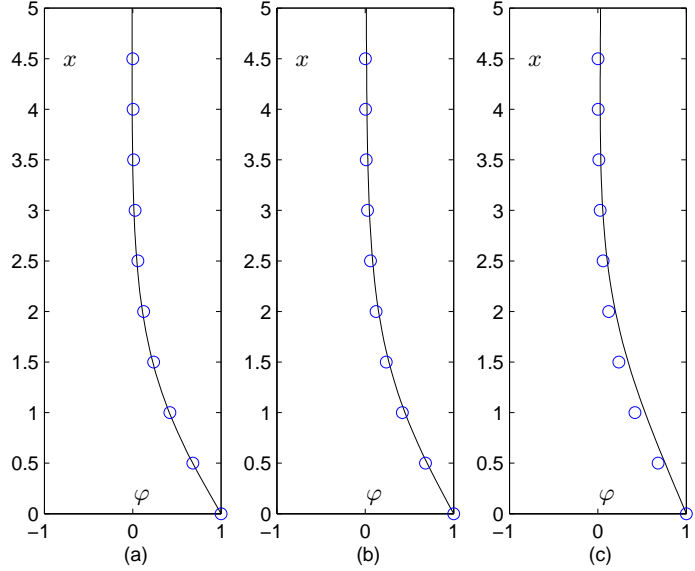}
\caption{\label{figdefdar}Mode shape at the critical load for a long beam. (a) $\eta=0$, (b) $\eta=0.3$, (c) $\eta=1$. (---), numerical results for $ \ell=10$; (o), solution for the semi-infinite beam  using matched assymptotic expansions, Eq.(\ref{equffipc}). }
\end{center}
\end{figure}

The particular value of the  flutter instability threshold $q_c$ may be directly derived by assuming that $\omega$ is real in  Eq.(\ref{equomegap}). As all functions of $q$ in (\ref{equomegap}) are real this implies  that
  \be 
\label{equpc} 
\Ai'(-q_c)=0 ; \ q_c  \simeq 1.02.
 \ee
 The real part of the  frequency at the instability threshold is then   derived using (\ref{equomegap} 
) and (\ref{equpc} ) as
  \be 
\label{equomegac} 
\omega_c^2= \eta \frac{\Ai(-q_c)}{(1/3)-\ba{\Ai}(-q_c)}. 
 \ee
At $\eta=0$, the solution of \cite{triantafyllou1984a} for divergence  is recovered, both in terms of critical load and frequency.  
Using this approximate solution, flutter is found to  exist for all values of $\eta$, and the corresponding critical load does not vary with  $\eta$. In Fig.  \ref{figapproxdar}(a) this is compared with the  numerical solution for very long beams, $\ell=100$.
The approximate solution, equation (\ref{equpc}),  predicts a lower bound, $q_c=10.2$,  equal to that for $\eta = 0$.
The flutter frequency is very  well predicted, as may be seen in Fig.\ref{figapproxdar}(b). 
%Note that this approach may be improved by also lumping the rotary inertia of the inner domain at $x=0$. The critical load is then  found to  vary with $\eta$ in a manner closer to the numerical results for long beams. This is not presented here for the sake of clarity.   
At the instability threshold the modal shape $\ffi$ is real and reduces to 
 \be 
\label{equffipc} 
\ffi(x)=  \int_0^x Ai(s-p) \  ds -\frac{1}{3}.
 \ee
This is compared, Fig.\ref{figdefdar}, with the computed mode shapes at $\ell=10$. For $\eta=0$, Fig.\ref{figdefdar}(a), Eq. (\ref{equffipc}) is actually the exact solution for a semi-infinite beam, which derives from  the results of \cite{triantafyllou1984a}.  For  $\eta=0.3$ and even  $\eta=1$,  it is remarkable that Eq.  (\ref{equffipc}) stills gives a very good approximation of the mode shape at the flutter threshold, Figs. \ref{figdefdar}(b) and (c). 

%%%%%%%%%%%%%%%%%%%%%%

\subsection{A model  based on  local wave stability}

Following  \cite{doare2002a}, we consider now a criterion based on  the local characteristics of waves, at a given position $x$.  From Eq. (\ref{equsuivadim2}), the dispersion relation  is 
\be
k^4-(q-x)k^2-\omega^2=0
\ee
where $k$ is the wavenumber  and $\omega$ is the frequency of the wave. 
All points such that $x<q$ bear unstable waves.  Beams of length $\ell$ larger than $q$ are expected to have a behaviour not affected by length as  waves are damped  above $x=q$ 

In dimensional variables this allows  us to define a the length $L_N$ for neutral stability,
\be
L_N=\frac{P}{mg},
\ee 
above which the medium only bears stable waves. 
In Eq. (1) this is  the point where the local tension goes through zero. Considering  that the part of the beam above this, $Z>L_N$, plays a negligible role in the instability, we may approximate the critical load for the {\em semi-infinite} beam by that for a {\em finite} beam of length $L_N$:  a beam of length $L_N$, without gravity, has a  dimensional critical load 
\be
P_c=\frac{EI}{L_N^2}p_c. 
\ee
where $p_c$ is the dimensionless critical load  for short beams. 
As the length $L_N$ varies  with the  load $P$, this results in 
\be
P_c^3=EI (mg)^2 p_c. 
\ee
Using  now dimensionless variables pertaining to the case of long beams, this reads
\be
\label{equlocalflutt}
q_c=p_c^{1/3};
\ee
similarly, we have
\be
\label{equlocalomegac}
\omega_c  =\frac{\Omega_c}{p_c^{2/3}}.
\ee

\begin{figure}
\begin{center}
\psfrag{psl}{$\ell$}
\psfrag{psqc}{$q_c$}
\psfrag{pseta}{$\eta$}
\psfrag{psa}{(a)}
\psfrag{psb}{(b)}
 \centerline{
\includegraphics[width=7cm]{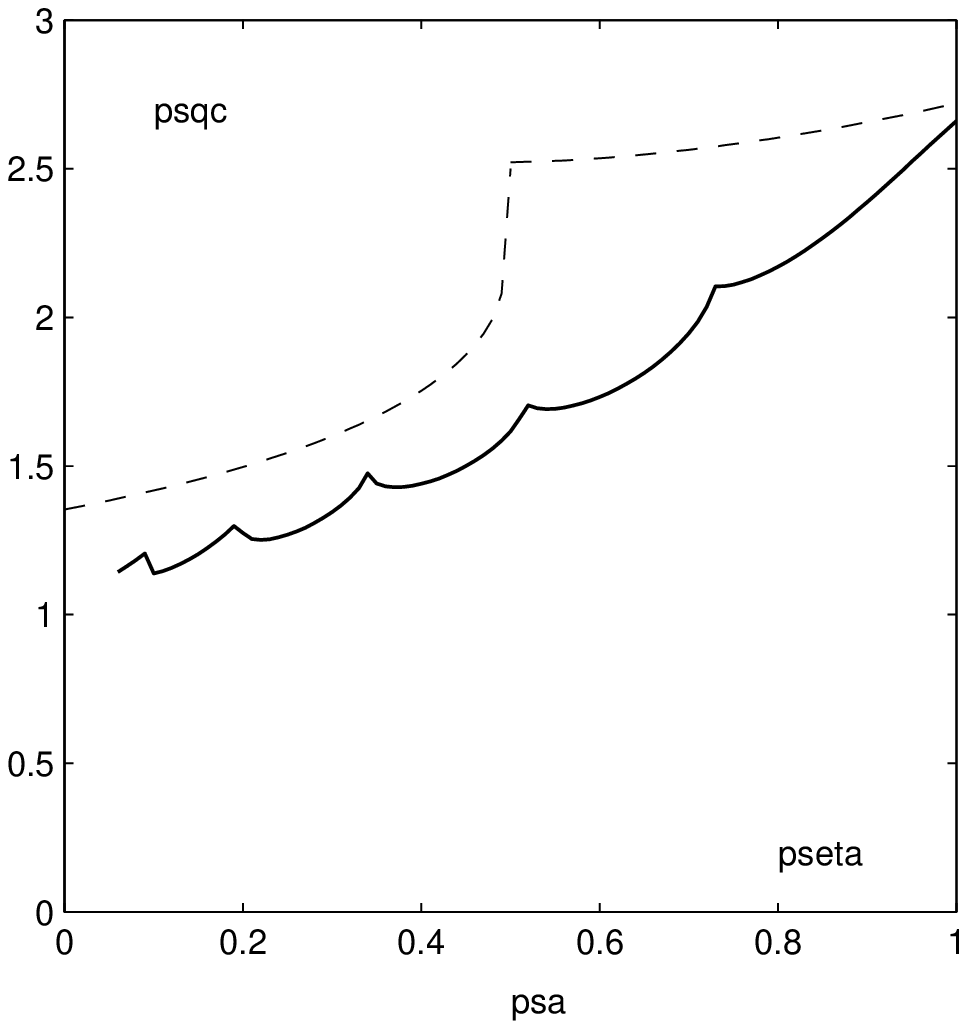}
\includegraphics[width=7cm]{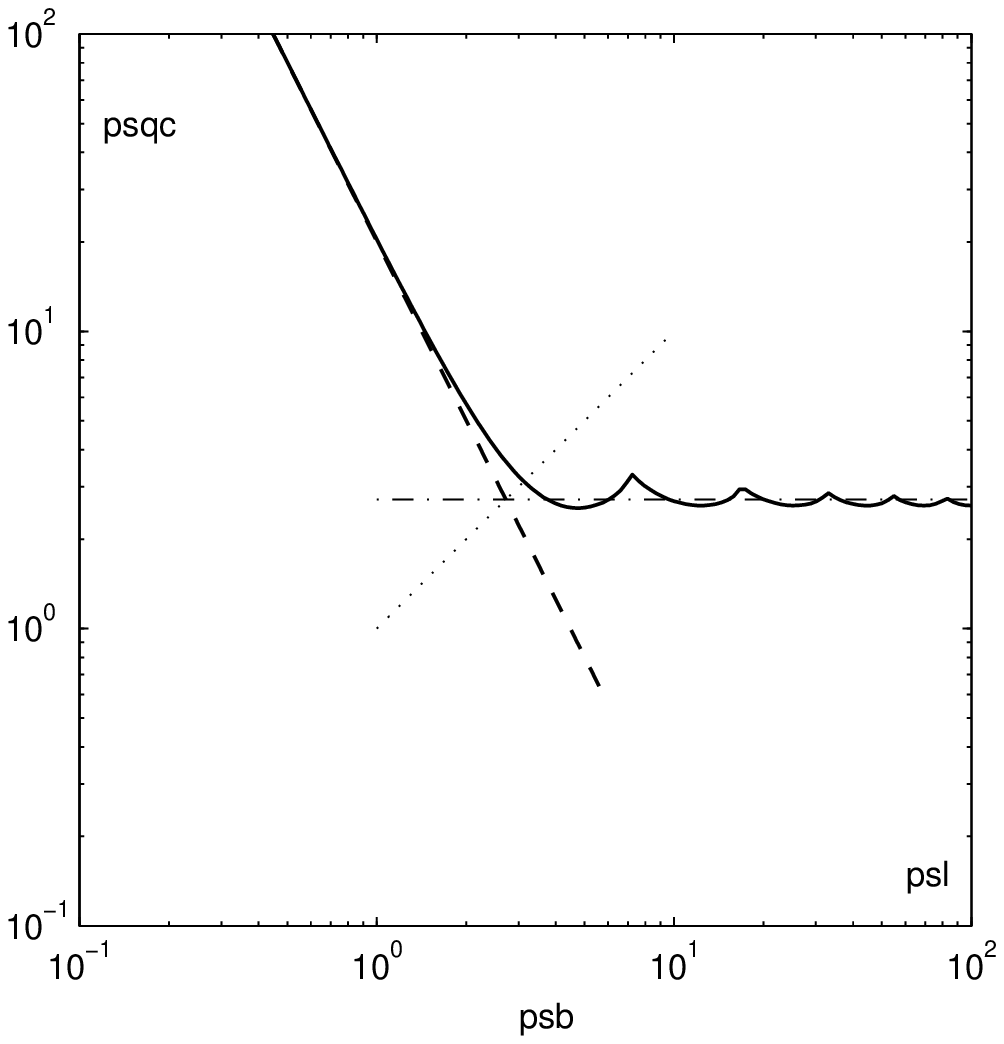}}
\caption{\label{figeffetlonglocal}Models for the critical load  based on  local wave stability, (a)  critical load (---) numerical results for $\ell=100$; (- -), model  based on  local wave stability, Eq. (\ref{equlocalflutt}) and (\ref{equlocalomegac}). (b) Effect of length $\eta=1$. (---), computations of Section \ref{sec:lengtheffect}; (- -), short beam model; (...), transition line; (-.-), long beam model. }
\end{center}
\end{figure}

In Fig.   \ref{figeffetlonglocal}(a)   this is plotted for  comparison with numerical results for the very long beam. The order of magnitude of the critical load is well recovered. Yet, divergence is predicted for $\eta<0.5$, instead of flutter. Note also that the critical load at $\eta=1$ is  very well predicted.  
Figure \ref{figeffetlonglocal}(b) summarizes this approach: the limit line $\ell=q$ is found to be a good approximation of the transition between the behaviour of short beams, that depend on length, and that of long beams, that does not. For short  beams the approximation of Eq. (\ref{equshort}) applies. For long beams Eq. (\ref{equlocalflutt}) is a good approximation.

%%%%%%%%%%%%%%%%%%%%%%%%%%%%%%%%%%%%%%%%%%%%%%%%%%%%%%%%%%%%%%%%%

\section{\label{sec:discussion}Discussion}

%%%%%%%%%%%%%%%%%%%%%%

\subsection{Domains of instability}

In the results presented in this paper  two regions can be indentified in a beam tensionned by a load such as gravity. In the part near the free end, $x\ll 1$ in our dimensionless variables, that is $Z \ll L_g=(EI/Mg)^{1/3}$, the stiffness that opposes lateral displacement is dominantly that of the flexural rigidity. Conversely for $x \gg 1$ the stiffness is due to the tension that results from gravity. 

For a given beam, if its length $L$  is much smaller than $L_g$, or $\ell \ll 1$, gravity effects can be neglected everywhere. In terms of stability the  behaviour is that of short beams in our denomination. The upper boundary condition, here clamping, plays a crucial role, and the instabilty may be divergence or flutter. If the length is much larger than $L_g$, so that $\ell \gg 1$, gravity effects dominate in all the upper part. This results in a confinement of the instabilty in the lower part of the beam, as it appeared in all the mode shapes. The upper boundary condition plays a lesser role. It is remarkable that  the characteristics of  the instability  of these long beams seem to be  much simpler than that of short beams: flutter prevails in all cases, and  the critical load, the frequency and the mode shape have very simple evolutions. 

For long beams flutter is reached when the load $q$ is of order $1$. In that range of loading the neutral point where tension vanishes, at $x=q$, is located near the limit between the two regions, $x=1$. The critical load is then such that, in dimensionless variables
\be
P \simeq (EI)^{1/3} \ (mg)^{2/3},
\ee
which shows the balance between the loading $P$  and the two stiffnesses of the systems, $EI$ for the flexural rigidity and $mg$ for the tension induced by gravity. 

It should be recalled here that damping was not considered in our calculations. Damping may significantly modify the critical value for flutter instability even when small \citep{langthjem2000a}, and has been found to have both a stabilizing and destabilising effects. In some cases, such as that of Beck's column, it has been found that addition of damping destabilizes the system and changes the instability type from flutter to buckling. In gyroscopic systems such as fluid conveying-pipes or plates in axial flows, damping was found to destabilise neutral waves at zero frequency \cite{doare2010,doare2014}, which may also cause the instability to change from flutter to buckling in finite length systems. The generalisation to the present case of a semi-infinite system tensionned by gravity needs to be addressed to see how damping could alter the buckcling/flutter stability maps, and if it prevents flutter to prevail for all values of the follower force coefficient $\eta$.

%%%%%%%%%%%%%%%%%%%%%%

\subsection{Relation to flow-induced instabilities}

The equivalence between the loading produced by the flow along a structure and a follower force exerted at its free  end is discussed in \cite{paidoussis1998a,paidoussis2003a}. Depending on the geometry of the  problem, the load maybe either a fully follower load (fluid-conveying pipe) or partially a follower load (cylinder with axial flow). In the latter case, the shape of the downstream free end strongly influences the value of $\eta$.

For systems  mainly tensionned by gravity, such as  hanging fluid-conveying pipes  or  hanging ribbons  under axial flow   \citep{doare2002a,lemaitre2005a} the equations of motion  can be put in the   common dimensionless form  
\be
\label{equsuivadim3}
 \frac{\partial^4 y}{\partial x^4}+ \frac{\partial}{\partial x}\left[(v^2-x)\frac{\partial y}{\partial x}\right]+ 2 v \sqrt{\beta}\frac{\partial^2 y}{  \partial x \partial \tau }+ \frac{\partial^2 y}{\partial \tau^2}=0,
 \ee
where the dimensionless variables have been defined using the length $L_g$, as we did in the present paper. Here $v$ is a dimensionless flow velocity and $\beta$ expresses the proportion of the fluid mass in the total mass. The boundary conditions at the lower end read
\be
\label{equbc3}
 \frac{\partial^2 y}{\partial x^2}(0)=0  ;  \  \frac{\partial^3 y}{\partial x^3}(0)=0, 
 \ee
which expresses the case of a purely follower load. Eq.  (\ref{equsuivadim3})  is identical to (\ref{equsuivadim2}), except for the gyroscopic term  which is dependent on $\beta$. The stability diagram for the case of a long hanging pipe has been analysed by \cite{doare2002a} and for  long hanging ribbons by \cite{lemaitre2005a}. The critical velocity $v_c$ was found to depend on the length in a manner very  similar to that found in this paper. For very long pipes flutter  arises at a critical velocity $v_c$ that depends only on  $\beta$. Defining $q_c=v^2$, and in the particular limit of $\beta=0$, the results of \cite{doare2002a}  should converge to that found in this paper for $\eta=1$. This is not exactly the case when the results of \cite{doare2002a} are used, as there calculations have been done for small but non vanishing  values of  $\beta$. It is a well known feature that the case $\beta=0$ is only found as the limit of very small values ($10^{-3}$) of $\beta$  \citep{paidoussis1998a}.

If the tensioning load is not gravity but a flow-induced friction on the slender structure, the form of the problem is changed in several ways  \citep{paidoussis2003a}. First, the tensioning load  increases with the flow velocity, which does not allow one  to define an  equivalent to  the length $L_g$, as for gravity. The  reference length that can  then be defined scales the flow-induced forces that are proportionnal to volume (added stiffness forces) and those that are proportionnal to surface (friction). 
Second, the friction load is not constant in direction, as is gravity, but acts tangentially to the  instantaneous position of  the beam. Third, in the case of axial flow outside of a structure, the flow may also induce forces perpendicular   to the  instantaneous position of the beam, in the form of a transverse drag. Moreover, the boundary condition at the free end include other terms that depend on the local geometry of the flow.
If all time dependent terms are neglected, to determine the static stability of the beam under flow, the equation defining the deflection reads \citep{triantafyllou1984a}
\be
\label{equsuivadim4}
 \frac{\partial^4 y}{\partial s^4}+ v^2\frac{\partial}{\partial s}\left[(1-s)\frac{\partial y}{\partial s}\right]=0. 
 \ee
Upon defining $x=v^{2/3}s $ and $q=v^{2/3}$, the equation for the static behaviour of the beam, Eq. (\ref{equstat}), is recovered. If the end-condition sustains a force only in the beam axis, like in reference \cite{Naguleswaran2004a}, the results at $\eta=0$ are recovered. For the analysis of the dynamic instability not only inertial terms need to be considered, as in the case of  this paper. A gyroscopic term, as in Eq. (\ref{equsuivadim3}), and a damping term appear, further complicating the behaviour of  the system. In the analysis of long towed cylinder \cite{delangre2007a} edge flutter was also found, confirming the generality of the results presented here in a generic case.  

%%%%%%%%%%%%%%%%%%%%%%%%%%%%%%%%%%%%%%%%%%%%%%%%%%%%%%%%%%%%%%%%%

\bibliographystyle{plain}

%%%%%%%%%%%%%%%%%%%%%%%%%%%%%%%%%%%%%%%%%%%%%%%%%%%%%%%%%%%%%%%%%

\end{document}